\documentclass{emulateapj}
\usepackage{amssymb}

\def\h0 {$h_0$=71 km s$^{-1}$ Mpc$^{-1}$}

\newcommand{\be}{\begin{equation}}
\newcommand{\ee}{\end{equation}}
\newcommand{\ce}{\ifmmode {\cal E} \else ${\cal E}$\ \fi}
\newcommand{\kms}{\ifmmode {\rm km\ s}^{-1} \else km s$^{-1}$\ \fi}
\newcommand{\tes}{\ifmmode \tau_{\rm es} \else $\tau_{\rm es}$\ \fi}
\newcommand{\tk}{\ifmmode \tau_{\rm K} \else $\tau_{\rm K}$\ \fi}
\newcommand{\vfwhm}{\ifmmode V_{\mbox{\tiny FWHM}} \else
            $V_{\mbox{\tiny FWHM}}$\fi}
\newcommand{\msun}{\ifmmode M_{\odot} \else $M_{\odot}$\ \fi}
\newcommand{\afe}{\ifmmode {\mathcal A_{\rm Fe}} \else${\mathcal A_{\rm Fe}}$\ \fi}

\newcommand{\lb}{\ifmmode L_{\rm Bol} \else $L_{\rm Bol}$\ \fi}
\newcommand{\ledd}{\ifmmode L_{\rm Edd} \else $L_{\rm Edd}$\ \fi}
\newcommand{\lx}{\ifmmode L_{\rm 2-10keV} \else  $L_{\rm 2-10keV}$\ \fi}
\newcommand{\hb}{\ifmmode H\beta \else H$\beta$\ \fi}
\newcommand{\mbh}{\ifmmode M_{\rm BH}  \else $M_{\rm BH}$\ \fi}
\newcommand{\lv}{\ifmmode \lambda L_{\lambda}(5100\AA) \else $\lambda L_{\lambda}(5100\AA)$\ \fi}

\def\ariel5{{\it Ariel 5}\ }

\def\heao1{{\it HEAO~1}\ }

\shorttitle{Universal scaling relation for black holes}
\shortauthors{Zhou, Yuan, Pan \& Liu}

\begin{document}
\title{Universal scaling of the 3:2 twin-peak quasi-periodic oscillation frequencies with black hole mass and spin revisited}

\author{
Xin-Lin Zhou$^{1}$, Weimin Yuan$^{1}$, Hai-Wu Pan$^{1}$, Zhu
Liu$^{1}$}

\affil{
$^1$ National Astronomical
Observatories, Chinese Academy of Sciences, 20A Datun Road, Chaoyang District, Beijing, 100012, China; zhouxl@nao.cas.cn \\
}
\email{zhouxl@nao.cas.cn}


\begin{abstract}
We discuss further observational support of an idea formulated
 a decade ago by Abramowicz, Klu{\'z}niak, McClintock and Remillard.
They demonstrated
 that the 3:2 pairs of frequencies of the twin-peak black hole (BH) high-frequency
 quasi-periodic oscillations (QPOs) scale inversely with the BH masses and that the
 scaling covers the entire range from stellar to supermassive BHs.
 For this reason, they believed that the QPOs may be used for
accurate measurements of masses and spins of BHs.

\end{abstract}


\section{Introduction}
Multiple quasi-periodic oscillations (QPOs) are often found in the
power spectrum of X-ray variability in X-ray binaries.
High-frequency QPOs (HFQPOs; $40-450$ Hz) have been detected in a
few X-ray binaries (Remillard 2004). These HFQPOs may correspond to
the frequency at the innermost stable circular orbit for
Schwarzschild black hole (BH) with masses of $5-15$ $M_\odot$. These
QPO frequencies do not change significantly despite sizable changes
in the X-ray luminosity. This suggests that the HFQPO frequencies
are primarily dependent on the mass and spin of the BH (Remillard \&
McClintock 2006). Thus HFQPOs are likely to offer the most reliable
measurement of spin once the BH mass is known and correct model is
applied (e.g., Abramowicz \& Klu\'zniak 2001; Rezzolla et al. 2003).

QPOs have been detected in a few objects, e.g., the ultra-luminous
X-ray sources NGC 5408 X-1 (Strohmayer et al. 2007), M82 X-1
(Strohmayer \& Mushotzky 2003; Feng \& Kaaret 2007;
Caballero-Garc\'ia et al. 2013; Pasham et al. 2014),  the Seyfert
galaxy RE J1034$+$396 (Gierli\'nski et al. 2008; Alston et al.
2014a), and the tidal disruption event Swift J1644$+$57 (Reis et al.
2012) . Are these QPOs the analogies of HFQPOs? Can the BH spin play
a role? Here we study the correlation between the QPO frequency and
BH mass for these objects. We find a universal scaling relation from
Galactic X-ray binaries to active galactic nuclei.

There is a simple theoretical reason for the inverse mass
 scaling of the BH QPOs frequencies. Indeed, strong
 gravity has a characteristic lengthscale $R_{\rm G} = 2GM/c^2$,
 which formally corresponds to a characteristic frequency
 $\nu_{\rm G} = c/R_{\rm G} = c^3/(2GM) \sim 1/M$. If the
 observed BH QPO frequencies reflect some orbital
 (geodesic) frequencies, then one would expect that
 $\nu_{\rm QPO} = [c^3/(2GM)]f(x,a)$, where $x = r/R_{\rm G}$ is
 a dimensionless radius at which the oscillation occurs, and $a$
 is the dimensionless spin. Abramowicz \& Klu{\'z}niak (2001)
 were the first to  noticed¡ªin the observational data
 --- that the observed twin-peak QPO frequencies are in a 3:2
 ratio,\footnote{Titarchuk (2002) wrote about the chronology of
 these observational discoveries: ``Recently Abramowicz et al.
 (2003) drew the attention of the community to the possible
 presence of the 3:2 resonance in QPOs for the neutron star
 sources. A similar effect was also pointed out earlier
 by Abramowicz \& Klu{\'z}niak (2001) and Remillard et al.
 (2002) for the BH sources." Note also that neither
 in Strohmayer (2001a), nor in Strohmayer (2001b), the ratio
 3:2 is mentioned. Although Strohmayer discovered the
 twin-peak BH QPOs, he did not notice their
 3:2 frequency ratio.} and argued that the coherence of QPO,
 discussed above, suggests a {\it resonance}. In this case,
 $x = x_{\rm res}(a)$ is fixed by the very nature of the
 resonant oscillations that are behind the observed 3:2
 ratio of the twin-peak QPO frequencies, and
 $\nu_{\rm QPO} = [c^3/(2GM)]f(x_{\rm res}(a), a)
 = [c^3/(2GM)] F(a)$, with $F(a)$ being known for any specific
 type of resonance. Thus, the ``3:2 QPO frequencies" depend on
 mass and spin through a known relation,
 $\nu_{\rm QPO} = \nu_{\rm QPO}(M, a)$, and
 measuring $\nu_{\rm QPO}$ may be an accurate tool for measuring
 mass and spin. This idea, first applied by Abramowicz and
 Klu{\'z}niak (2001) for the Galactic BH QPOs was later
 employed by numerous authors, e.g. T{\"o}r{\"o}k et al. (2005),
 also for SgrA* T{\"o}r{\"o}k (2005), and more recently
 by Abramowicz \& Liu (2012) for the Swift J164449.3+573451
 supermassive BH.

 \par Our Letter is based on the paper by Abramowicz et al. (2004).
 They stressed the importance of the universal $\nu \sim 1/M$
 scaling and suggested that it may be used for a definite mass
 measurement of the hypothetical intermediate BH, if the
 twin-peak 3:2 QPOs would be discovered for such sources.
 Indeed, Pasham et al. (2014) have recently discovered
 a 3:2 twin-peak QPO in the ultra-luminous source M82 and claimed,
 following the Abramowicz et al. (2004) suggestion, that this is a
 $400 M_{\odot}$ BH.
Throughout this work, we assume the concordance cosmology of \h0 ,
$\Omega_{m}=0.27$, and $\Omega_{\Lambda}=0.73$.

\section{Objects with (HF)QPO detection}
Table 1 lists the objects with the (HF)QPO detections. Four BH X-ray
binaries, GRO J$1655-40$, XTE J$1550-64$, GRS $1915+105$, and H
$1743-322$ display a pair of HFQPOs with a 3:2 frequency ratio. The
frequencies of X-ray binaries are listed in Table 1 for the stronger
QPO that represent $2\times \nu_{0} $. The former three X-ray
binaries have well-measured dynamical BH mass. Literatures by
different authors gave slightly different values for BH mass.
Recently, Reid et al. (2014) measured a trigonometric parallax for
the X-ray binary GRS $1915+105$.  This gives a direct distance
estimate and a revised estimate for the BH mass of
$12.4^{+2.0}_{-1.8}$ $M_{\odot}$ . This is still in the range of
$10-18$ $M_{\odot}$  used in Remillard \& McClintock (2006).

QPOs have been detected in the ultra-luminous X-ray source M82 X-1
(Strohmayer \& Mushotzky 2003; Feng \& Kaaret 2007).
Caballero-Garc\'ia et al. (2013) reported on the detection of
($46\pm2$) mHz QPOs in the power density spectra of M82 X-1. A
comparison of the frequency of these QPOs with previous detections
supports the 3:2:1 frequency distribution. If it is true, this
implies a BH mass of $\sim10^4$ $M_{\odot}$ for M82 X-1. However,
Pasham et al. (2014) presented twin-peak (3:2 frequency ratio) X-ray
QPOs from M82 X-1 at frequencies of $3.32 \pm 0.06$ Hz and
$5.07\pm0.06$ Hz. They also estimated a BH mass of $415\pm63$
$M_{\odot}$ using the relativistic precession model.

The ultra-luminous X-ray source NGC 5408 X-1 shows a QPO at 20 mHz
(Strohmayer et al. 2007). They also found evidence for a second QPO
peak at 15 mHz.  The frequency ratio of this QPO pair is
inconsistent with 3:2 but is consistent with a 4:3 ratio. The BH
mass of this source is still under debate. Recently, Huang et al.
(2013) presented X-ray timing and spectral analysis to support a
large BH mass of $\sim10^5$ $M_{\odot}$. The X-ray light curve of
one 90 ks XMM-Newton observation of the Seyfert galaxy RE
J$1034+396$ shows an $\sim1$ hour X-ray periodicity (Gierli\'nski et
al. 2008). The QPO feature is detected in five low-flux/spectrally
harder observations (Alston et al. 2014a). This source is a narrow
line Seyfert 1 galaxy. The BH mass of $4^{+3}_{-1.5}$ $\times10^6$
$M_{\odot}$ for this object is derived from X-ray variability
amplitude (Zhou et al. 2010). This is broadly in agreement with the
value of $(1-4)$ $\times10^6$ $M_{\odot}$ derived from its bulge
stellar velocity dispersion (Bian \& Huang 2010).

Swift source J$164449.3+573451$ (Swift J$1644+57$) at redshift $z =
0.3543$ recently experienced a powerful outburst, most likely caused
by a tidal disruption of a star by the massive BH at the center of
the source. During the recent X-ray outburst, the source was similar
to a small-scale blazar (Bloom et al. 2011), with a relativistic jet
(Burrows et al. 2011). Reis et al. (2012) detected a firm QPO from
this object with a centroid frequency $\sim4.8$ mHz. It has been
suggested to be an intermediate-mass BH in Swift J$1644+57$
(Abramowicz \& Liu 2012; Shen \& Matzner 2014). Miller \& G\"ultekin
(2011) combined X-ray and radio observations of Swift J$1644+57$ to
constrain BH mass of $\sim(3.16^{+35.6}_{-2.9})$ $\times10^5$
$M_{\odot}$.

The QPOs have been detected from the infrared (Genzel et al. 2003),
X-ray (Aschenbach et al. 2004a) and radio bands (Miyoshi et al.
2011) in the Galactic Center Sgr A*. X-ray observations have
revealed a 3:2:1 frequency ratio for Sgr A* (Aschenbach et al.
2004a), although the significance of the QPOs is still questionable.
The dense gas cloud called G2 moving towards Sgr A* (Gillessen et
al. 2012) will produce strong flare at multiple bands and will test
the possible QPO emission in Sgr A*.

\section{Results and discussion}

Figure 1 shows the relationship between the QPO frequency and the BH
mass listed in Table 1. Three BH X-ray binaries, XTE J$1550-564$,
GRO J$1655-40$, and GRS $1915+105$, have well-measured dynamical BH
masses. The HFQPO frequencies from the three BH X-ray binaries
appear to scale inversely with mass. The relationship based on these
three stellar-mass BHs for the stronger $2\times$ $\nu_{0}$
frequency is (Remillard \& McClintock 2006),
\begin{equation}
 \nu (Hz) =1862(M_{\rm BH}/M_{\odot})^{-1}.
\end{equation}

We call this relationship the universal scaling relation. This
relation is plotted in Figure 1 as the solid line, and extrapolated
to the higher-mass range. If these HFQPOs are indeed general
relativity oscillations, then the simple scaling with BH mass may
suggest that three stellar BHs have similar values of the
dimensionless spin parameter, $a$ (Remillard \& McClintock 2006).
The spins for the three BHs have been measured from the X-ray
continuum fitting (McClintock et al. 2011) and X-ray reflection
method (Reynolds 2014). GRS $1915+105$ (McClintock et al. 2006; Blum
et al. 2009) has an extremely high spin value ($a\sim 0.99$), while
GRO J$1655-40$ (Reis et al. 2009) and XTE J$1550-564$ (Steiner et
al. 2011) has a moderate spin value.

It is suggested that an HFQPO with the harmonic pairs of frequencies
in a $3:2$ or $3:1$ ratio arises from some type of resonance
mechanism (Abramowicz \& Klu\'zniak  2001; Remillard et al. 2002;
Abramowicz et al. 2004). The results from the HFQPOs using resonance
model (Abramowicz \& Klu\'zniak 2001) or another model (Rezzolla et
al. 2003) also suggest moderate or high spin values for the three
stellar BHs.

Based on the resonance model, for a given spin value, the
correlation between HFQPO frequency and BH mass is determined. The
relationship derived from a model of $3:2$ resonance and $a=0.996$
is (Aschenbach 2004b),
\begin{equation}
\nu (Hz) =2030.8(M_{\rm BH}/M_{\odot})^{-1}.
\end{equation}
As a comparison, we also show the relationship derived from a model
of $3:1$ resonance and $a=0.996$ (Aschenbach 2004b),
\begin{equation}
 \nu (Hz) =3068.9(M_{\rm BH}/M_{\odot})^{-1}.
\end{equation}
As shown in Figure 1, they are quite close to the universal scaling
relation.

Swift J1644$+$57 is a tidal disruption of a star by a dormant BH.
The BH mass estimated from radio observations (Miller \&  G\"ultekin
2011) for this object well follows the universal scaling relation,
although with a large error bar (see Figure 1). Swift J1644$+$57
shows evidence for the relativistic jet activity (Shao et al. 2011),
a probable two-component jet (Liu et al. 2012). The 4.8 mHz QPO
could be associated with a second, narrower jet periodically
sweeping the observer line of sight (Wang et al. 2014). If the 4.8
mHz QPO is due to the same mechanism as what produces the HFQPOs in
Galactic X-ray binaries, a very realistic possibility is that the BH
in Swift J1644$+$57 has a mass closer to the most likely mass of
$3\times10^{5}$ $M_\odot$ (Abramowicz \& Liu 2012), and has a high
spin (Wang et al. 2014). The rapid spin of a BH may be a necessary
condition to generate a relativistic jet.

It is very interesting that a few objects well follow the universal
scaling relation. Note that RE J1034$+$396 is a Seyfert galaxy, NGC
5408 X-1 and M82 X-1 are ultra-luminous X-ray sources, and Swift
J1644$+$57 is a tidal disruption of a star by a dormant BH. They are
different types of BHs and have different formation histories. All
these BHs follow the universal scaling relation defined from the
three Galactic X-ray binaries. This suggests that these BHs have
similar spin values, which are very likely moderate or high spin
values.

The spins of about a dozen of stellar BHs and about 20 supermassive
BHs have been measured from the X-ray reflection method (Reynolds
2014). Many objects have moderate or high spin parameters.  This
implies that the cosmic processes to spin up BHs are very effective.
Kerr BHs may be common at all scales (Volonteri et al. 2005). The
narrow-Line Seyfert 1 galaxy RE J$1043+396$ slightly deviates from
the universal scaling relation. If the BH mass of RE J1034$+$396 is
close to the most likely mass of $4\times10^{6}$ $M_{\odot}$ (Zhou
et al. 2010), it may have a lower spin than the stellar mass BHs.
This is in good agreement with the average spin measured from the
composite X-ray spectrum of narrow-line Seyfert 1 galaxies (Liu et
al. 2014). If the QPOs detected in Sgr A* are real, the well
measured dynamical BH mass of Sgr A* (Gillessen et al. 2009) follows
the 3:1 resonance line (see Figure 1). The BH in Sgr A* may have an
extreme spin parameter (Aschenbach 2004b).

Stellar BHs, such as BH X-ray binaries and microquasars, form via
stellar processes. Accretion and angular momentum extraction cannot
significantly change the spin parameter of a BH in a binary system
(King \& Kolb 1999; Zhang 2013). Thus the spin parameter of a
stellar BH reflects the value at formation.  For massive BHs, recent
studies link the evolution of BH spin to the distribution of
orientations of episodic accretion periods (e.g., Wang et al. 2009).
Most massive BHs would spin down if the orientations of the
accretion disks are isotropic; most massive BHs would spin up if the
distribution of the orientations is highly anisotropic or if
accretion is coherent for a long time (Dotti et al. 2013). The
formation history of intermediate-mass BH remains unknown. It is
likely that they are relic BHs in the early universe, with the spin
obtained from the early formation process. If we accepted that Kerr
BHs are common at all scales, we expect that more objects with HFQPO
detection will follow the universal scaling relation in the future
(e.g., Alston et al. 2014b).

There is still no consensus on the {\it specific} oscillatory
 mechanism of the high-frequency, twin-peak BH QPOs.
 Two fundamental facts seem to be well-established, however,
 (1) These oscillations are characteristic of strong gravity.
 This follows from the $\nu \sim 1/M$ scaling first reported
 for the BH binaries by Remillard et al. (2002)
 (see also the Remillard \&  McClintock 2006 review article) and later generalized
 by Abramowicz et al. (2004) for all astrophysical BHs,
 including supermassive and (still hypothetical)
 intermediate-mass BHs; (2) These oscillations are
 characteristic of a non-linear resonance. This follows
 from the fact that their frequencies have a universal
 3:2 ratio, as first noticed, and stressed, by Abramowicz
 \& Klu{\'z}niak (2001).

\acknowledgments We thank the referee for very helpful comments that
improved the manuscript significantly. We thank L. J. Gou and S. N.
Zhang for reading the manuscript and giving helpful comments. This
work is supported by the NSFC grant 11033007, the Strategic Priority
Research Program ¡°The Emergence of Cosmological Structures¡± of the
Chinese Academy of Sciences, grant No. XDB09000000 and the XTP
project XDA04060604, EP project XDA04061100.

\clearpage

\begin{figure}
\includegraphics[width=11 cm, angle=270]{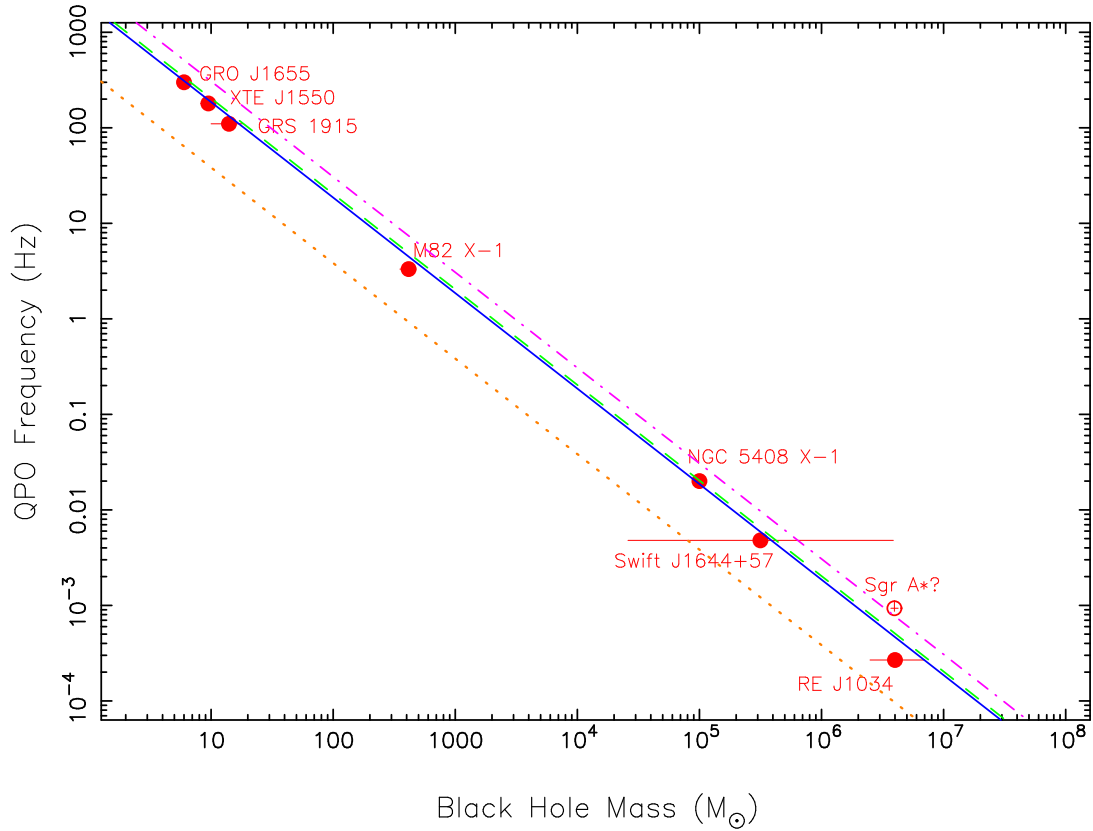}\label{fig1}
\caption  { Updated version of Figure 1 from Abramowicz et al.
(2004); see also Figure 2 in T{\"o}r{\"o}k (2005).
 In particular, scales and lines are similar and have the
 same meaning. Here, we merely add more data points which in 2004
 were not available.
Relationship between the quasi-periodic oscillation (QPO) frequency
and the black hole mass. Three X-ray binaries, XTE J1550$-$564, GRO
J1655$-$40, and GRS 1915$+$105, display a pair of high-frequency
QPOs with a 3:2 frequency ratio. The frequencies of X-ray binaries
are plotted for the stronger QPO that represent $2 \times \nu_{0}$.
The solid line denotes the relation, $\nu$ (Hz) = 1862$(M_{\rm
BH}/M_{\odot})^{-1}$ derived from three X-ray binaries (Remillard \&
McClintock 2006). The long-dashed line denotes the relation, $\nu$
(Hz) =2030.8$(M_{\rm BH}/M_{\odot})^{-1}$ for a model of 3:2
resonance and spin parameter $a=0.996$; the dotted-dashed line
denotes the relation, $\nu$ (Hz) =3068.9$(M_{\rm
BH}/M_{\odot})^{-1}$ for a model of 3:1 resonance and spin parameter
$a=0.996$; the dotted line denotes the Kepler frequency  for a
non-spinning Schwarzschild black hole at the innermost stable
circular orbit. The QPO reported in the active galactic nucleus REJ
1034+396, the ultra-luminous X-ray source NGC 5408 X-1, and the
tidal disruption event Swift J1644$+$57 is plotted. The observed QPO
frequencies in the ultra-luminous X-ray source  M82 X-1 display 3:2
ratio (Pasham et al. 2014). The frequency for this object also
represents $2 \times \nu_{0}$. The Galactic Center Sgr A* follows
the 3:1 resonance line, although the significance of QPOs in Sgr A*
is still under debate. See the text for details. }
\end{figure}

\begin{table*}
\caption{Lists of objects with (HF)QPO detection} \vglue 0.1cm
\label{tab:QPO} \centering
\begin{tabular}{l c c c c c c c}     
\hline
Source Name  & frequency   & 2$\nu_0$ & Ref1. &  $M_{\rm BH}$ &   Method   & Ref2.  \\
            &    (Hz)       &      &     &  ($M_{\odot}$)     &            &     \\  \hline
GRO J$1655-40$ & $300\pm10$ &  Yes &  1   &   $6.3\pm0.3$ &  D  &  9      \\
XTE J$1550-564$ & $180\pm8$ &  Yes &  1   & $9.6\pm1.2$  & D  &    9     \\
GRS $1915+105$  & $113\pm5$ &  Yes &  2   &  $12.4^{+2.0}_{-1.8}$& D  &  10      \\
H $1743-322$   & $163^{+5}_{-2}$ & Yes &  3   &    ... & ... & ... \\
M82 X$-1$        & $3.32\pm0.06$   & Yes &  4   & $415\pm63$   & P     & 4   \\
NGC 5408 X$-1$    & $(20.2\pm0.3)$E-3 & No & 5  & $(1\pm0.3)$E5  & T     & 11  \\
Swift J$1644+57$  & $(4.8\pm0.3)$E-3  & No & 6 &
$(3.16^{+35.6}_{-2.9})$E5  & R  &  12 \\
RE J$1034+396$    & $(2.68\pm0.1)$E-4 & No & 7  & $(4^{+3}_{-1.5})$E6  & T & 13 \\
Sgr A*            & $(1.07\pm0.05)$E-3 & Yes & 8 & $(4.31\pm0.42)$E6
&
D & 14\\
\hline
\end{tabular}
\begin{minipage}{\linewidth}
Ref1. $-$1: Remillard et al. (2002); 2:  Remillard (2004); 3: Homan
et al. (2005); 4: Pasham et al. (2014);  5: Strohmayer et al.
(2007); 6: Reis et al. (2012); 7: Gierli\'nski et al. (2008); 8:
Aschenbach et al. (2004a). Ref2. $-$9: Remillard \& McClintock
(2006); 10: Reid et al. (2014); 11: Huang et al. (2013); 12: Miller
\& G\"ultekin (2011); 13: Zhou et al. (2010); 14: Gillessen et al.
(2009).
 Method for black hole mass estimates. $-$D: Dynamical
method; P: relativistic precession model; T: X-ray timing; R: radio
observations.
\end{minipage}
\end{table*}

\end{document}